\documentclass[preprint]{iucr}              
\ifPDF
  \RequirePackage{hyperref}
  \PassOptionsToPackage{pdftex,bookmarksopen,bookmarksnumbered}{hyperref}
\fi
\usepackage{amssymb}
\usepackage{xcolor}
\usepackage{graphicx}
\usepackage{pdfpages}

\newcommand\PTCDIC{PTCDI-C\textsubscript{8} }
\newcommand\Alq{Alq3 }

     \journalcode{S}              

\begin{document}                  



\title{Closing the loop: Autonomous experiments enabled by machine-learning-based online data analysis in synchrotron beamline environments}


\cauthor[a]{Linus}{Pithan}{linus.pithan@uni-tuebingen.de}{}
\author[a]{Vladimir}{Starostin}
\author[b]{David}{Mareček}
\author[c]{Lukas}{Petersdorf}
\author[a]{Constantin}{Völter}
\author[a]{Valentin}{Munteanu}
\author[d]{Maciej}{Jankowski}
\author[d]{Oleg}{Konovalov}
\author[a]{Alexander}{Gerlach}
\author[a]{Alexander}{Hinderhofer}
\author[c]{Bridget}{Murphy}
\cauthor[b]{Stefan}{Kowarik}{stefan.kowarik@uni-graz.at}{}
\cauthor[a]{Frank}{Schreiber}{frank.schreiber@uni-tuebingen.de}{}

\aff[a]{Universität Tübingen, Inst. für Angewandte Physik, Auf der Morgenstelle 10, 72076 Tübingen,  \country{Germany}}
\aff[b]{Universität Graz, Physikalische und Theoretische Chemie, Heinrichstraße 28, 8010 Graz, \country{Austria} }
\aff[c]{Universität Kiel, Inst. für Experimentelle und Angewandte Physik, Leibnizstraße 19, 24118 Kiel, \country{Germany} }
\aff[d]{ESRF, The European Synchrotron, 71 avenue des Martyrs, CS 40220, 38043 Grenoble Cedex 9, \country{France} }







\maketitle                        


\newpage
\begin{abstract}
Recently, there has been significant interest in applying machine 
learning (ML) techniques to X-ray scattering experiments, which proves
to be a valuable tool for enhancing research that involves large or
rapidly generated datasets. ML allows for the automated interpretation
of experimental results, particularly those obtained from synchrotron or
neutron facilities. The speed at which ML models can process data
presents an important opportunity to establish a closed-loop feedback
system, enabling real-time decision-making based on online data
analysis. In this study, we describe the incorporation of ML into a
closed-loop workflow for X-ray reflectometry (XRR), using the growth of
organic thin films as an example. Our focus lies on the beamline integration of
ML-based online data analysis and closed-loop feedback. We present
solutions that provide an elementary data analysis in real time during the experiment without introducing the additional software dependencies in the beamline control software environment. Our data demonstrates the accuracy and robustness of ML
methods for analyzing XRR curves and Bragg reflections and its
autonomous control over a vacuum deposition setup.
\end{abstract}

\section{Introduction}\label{introduction}

X-ray user facilities are amongst the largest scientific data producers
in the world \cite{yanxon_artifact_2023}. While experiments performed
at these facilities cover an extensive range of multi-disciplinary sciences,
they typically share a common data generation pattern, namely precisely
positioning a specimen in the path of the X-ray beam and recording data
(e.g. radiation scattered by the sample) captured by dedicated detectors
(e.g. diffraction, imaging, spectroscopy) further downstream. Recent
advances in accelerator development such as fourth-generation
synchrotron light sources \cite{raimondi_esrf-ebs_2016} and innovative detector
technology (e.g. higher acquisition rates and larger area detector
dimensions) lead to continuously increasing data volume of these
datasets that are more and more difficult to handle. In fact, in order
to manage acquisition, analysis and storage of these, specific solutions
developed at the facilities \cite{guijarro_bliss_2018,allan_blueskys_2019,scicat} or in data-driven national and international collaborations, e.g. DAPHNE4NFDI  \cite{daphne4nfdi} or PaNOSC \cite{carboni_panosc_2022,panosc}, have been put in place, addressing these challenges.

Through the described technological advances and newly deployed
infrastructure at modern beamlines and facilities, the way experiments
are performed changes significantly towards more and more data-intense
and data-driven experiments, increasingly relying on machine learning
(ML) based approaches for data analysis \cite{chen_machine_2021, hinderhofer_machine_2023}.
While facility instruments used to be basically fully isolated systems, often posing complications for visiting facility users to
integrate sample-environment related equipment or data sources into the
beamline data acquisition system, this changes rapidly with the recent
developments in many aspects. This process is generally encouraged as the value of
freely available, complete, augmented and documented FAIR datasets \cite{wilkinson_fair_2016, scheffler_fair_2022} is recognized. These datasets are specifically valuable for machine learning activities and enable novel,
previously impossible experiments that fully utilize modern
infrastructure.

This paper introduces an approach for integrating real-time X-ray scattering and diffraction data analysis into a closed-loop control
system that actively adjusts sample environment parameters. This
approach unlocks exciting possibilities for conducting experiments using synchrotron radiation that
unveil new insights into the underlying physics. We demonstrate the seamless
integration of user-developed machine learning (ML) code with the
beamline control infrastructure, enabling real-time data analysis.
Additionally, focusing on reflectometry as a case study, we provide a
concise overview of a cutting-edge ML-based approach for predicting thin
film parameters in both single-layer and multilayer structures.

\section{Methods and Data}

Several recent publications highlight the use of ML in synchrotron and
neutron beamline environments \cite{noack_gaussian_2021, Yager_2023, beaucage_autonomous_2023, szymanski_adaptively_2023, kandel_demonstration_2023, suzuki_symmetry_2020,hinderhofer_machine_2023}. These developments are mainly initiated at large-scale facilities
or labs with privileged access rights that are necessary to integrate additional
software code into beamline software environments. However, often user-
and sample-specific ML techniques are needed in experiments. Therefore, this study explores potential approaches to deploy ML-based real-time analysis code at large scale facilities (X-ray and neutron sources) independent of the specific experiment or research facility. Where possible, we follow guidelines
based on community initiatives such as PANOSC \cite{panosc}
, EXPANDS \cite{expands}, DAPHNE4NFDI \cite{daphne4nfdi} or MLExchange \cite{zhao_mlexchange_2022}.

\subsection{Software environments and structural requirements}

Before discussing the architecture in detail, it is worth describing the
complete data acquisition and handling process on a conceptual level. In the
experiment described there are two main types of data generators: 1.
the area detectors as counters (specifically a Dectris Pilatus 300k and a MaxiPix 2x2 with a CdTe sensor)
controlled via LIMA \cite{petitdemange_lima_2018}, and 2. the motor
positioners. Experiment control is performed via BLISS \cite{guijarro_bliss_2018} which also handles the saving of the collected data in the NeXus file format (using HDF5) \cite{konnecke_nexus_2015} and triggers the
ingestion of data and meta-data into the facility data portal.

The incorporation of user-developed code into the software environment of the beamline through software interfaces is crucial for the presented experiment, but also for many future experiments yet to come.

Software interfaces that enable the embedding of user-developed code into the beamline software environment are crucial for the presented experiment and - in our view - most likely also for many future experiments yet to come. To ensure operational stability of instruments with
rapidly changing users there is an inherent dilemma from
the facility point of view, since allowing modifications to the
beamline's control software environment by one user puts the experiments of subsequent users at risk. Therefore, as a matter of best
practice, well isolated software environments for beamline control and
user code are crucial and one must find suitable methods to exchange
data between these environments and software processes. From a practical
point of view, such an approach also removes the burden of dealing with
incompatibilities in software dependencies (e.g., versions of specific Python
packages) that are used on both sides.

We identified different `hooks' that typically exist in a modern
beamline environment to achieve this inter-process communication between
the beamline control software and user-supplied data analysis code,
focusing on various levels of integration and portability between
different light sources.

Modern beamline control systems, such as BLISS available at the ESRF
\cite{guijarro_bliss_2018} and bluesky \cite{allan_blueskys_2019}, offer dedicated
frameworks to access data in the beamline environment e.g. through
``publishing'' the data through an integrated in-memory database or via
direct access to data- and event-streams produced by the acquisition
process. Evidently, these are hooks which could be used to integrate online data analysis
(see example in Supporting Information section SI-3), however these are
usually tightly coupled to a specific facility and typically introduce a
number of critical dependencies into the user supplied code.

Another relatively simple option for external users to handle and
test prior to the experiment, is to integrate on the level of the SCADA
(supervisory control and data acquisition) used at the respective
facility. The two most spread systems in this field are TANGO controls \cite{gotz_tango_2022, tango} or EPICS \cite{epics_website}. In the
context of this study, we evaluated the integration via TANGO that is
predominantly used at several European synchrotron facilities such as ESRF, DESY, SOLEIL or ALBA. Data transfer
from the beamline control system via the available SCADA system forms
part of the most common operations performed on the control system level
and thus does not induce any additional software dependencies for the
beamline control software environment.

As a third option, there is also the possibility to rely on workflow
engines offered by the facilities \cite{w_de_nolf_ewoks_2022}. However, care must be taken with
respect to the real-time capabilities of these systems, since they
inherently rely on queue systems (job schedulers for batch processing) and thus may introduce additional
delays under heavy load.

When it comes specifically to ML/AI pipelines/workflows, there is also
potential to rely on standardized solutions that specifically fit the
needs of handling larger ML models (e.g., NVIDIA Triton Inference
Server) and thus abstract even beyond the specificity of beamline
environments.

Evidently, combinations of all scenarios described above are possible
and were partially evaluated in this work.

\begin{figure}
\caption{Data flow during data collection, ML-based online data
analysis and closed loop feedback. Data is collected using Bliss,
transferred to online data analysis via tango and the analysis results are
fed into closed-loop operation as well as saved together with raw data
using Bliss again. Bottom left: experimental setup for molecular beam
deposition.}
\includegraphics[
  width=6.49426in,
  height=2.33172in,
  keepaspectratio]{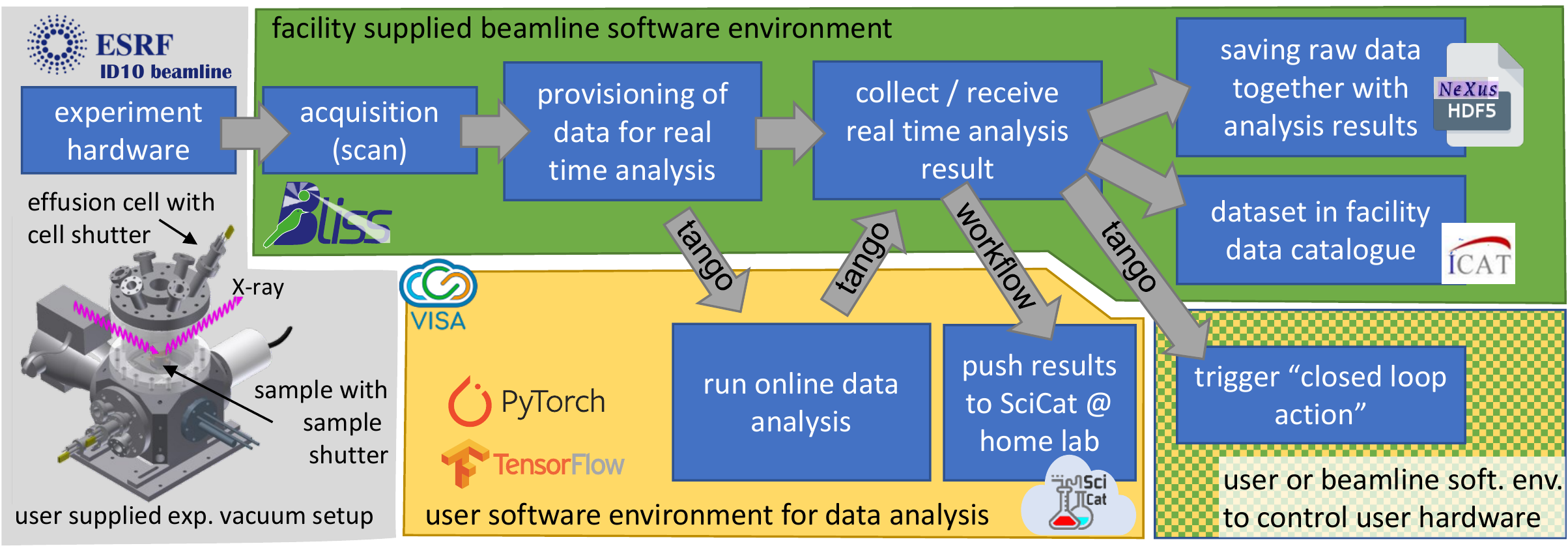}
\label{fig:dataflow}
\end{figure}

\subsection{Distributed online data analysis and closed-loop beamline
integration}

Splitting beamline control and ML based online data analysis in terms of
infrastructure makes sense also when looking at the different hardware
requirements of the two processes. While the beamline control process
links to beamline instrumentation, it is physically tied to hardware
that is available on the beamline itself. The ML data analysis process,
which requires a powerful GPU, can also run on edge computing nodes \cite{babu_deep_2022} or even in central computing facilities \cite{starostin_end--end_2022}. For more integrated, routine solutions
offered by the facilities themselves, edge computing is an attractive
option in this context. Prioritizing the user's requirements, it is advantageous for a facility to offer infrastructure that is
exclusively available to the particular user group performing the
experiment. Through the collaborative efforts in the context of PANOSC
the VISA system based on OpenStack \cite{visa_2023} has been developed which essentially fulfills the
needs described above.

Modern synchrotron beamlines are an ecosystem of distributed computing
resources on their own. Therefore, a simple choice to establish
communication between the beamline control software and the user-provided online data analysis resources is to rely on the beamline's
SCADA system (at ESRF: TANGO controls) for low dimensional data (sup.
inf. sections A and B). Streaming options, however, become inevitable for
high-rate, multidimensional data sources such as large area detector
images.

\begin{figure}
\caption{Illustration of timing and architecture of implemented
synchronous and asynchronous feedback loop. In a) data acquisition, ML
inference and feedback action follow each other strictly in time
(synchronous over distributed system) while in b) acquisition and data
analysis + feedback are separated in independent coroutines to decouple
independent processes. Further direct communication with the ML-model
using TANGO controls is illustrated in a) while in b) an intermediate
workflow system and an industrial inference server are used.}
\includegraphics[width=6.05963in,height=2.55152in,keepaspectratio]{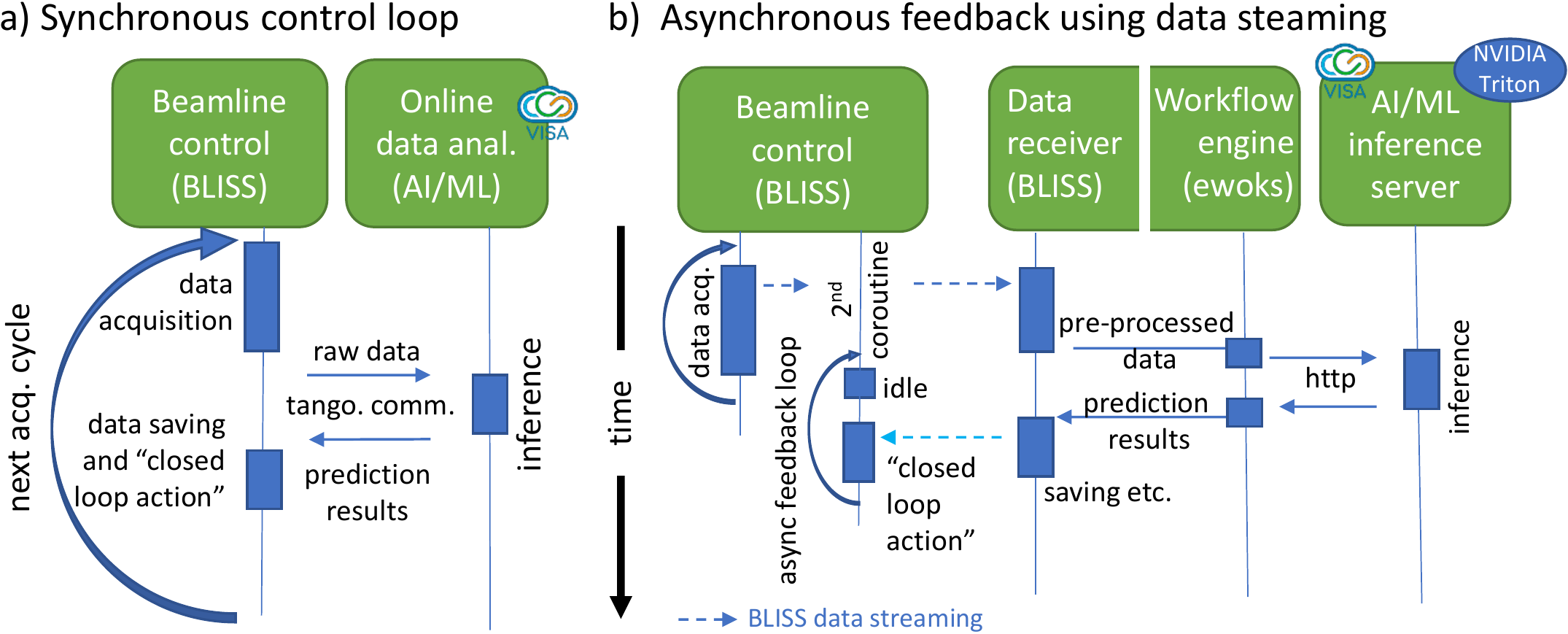}
\label{fig:architecture}
\end{figure}

In Fig. \ref{fig:architecture} we illustrate the layout of two independently tested software
configurations evaluated during the experiment presented. Fig. \ref{fig:architecture}a) shows
a fully \emph{synchronous} acquisition and online analysis scheme
relying on TANGO communication to transfer 1d data (binned ROI
intensities extracted from detector images, here done via LIMA \cite{petitdemange_lima_2018} to the online data analysis running
on a user controlled virtual machine provisioned via VISA where the ML
inference takes place. After processing ML-results are made available in the
main beamline control process and the `closed loop action' can be
triggered. Once completed, another acquisition starts (see pseudocode example
in Supporting Information section Si-2).

Fig. \ref{fig:architecture}b) shows a configuration that decouples the data acquisition and
analysis+feedback into independent processes (asynchronous feedback) to
maximize the data acquisition rates. Here, we use the streaming capabilities
of BLISS to transfer reflectometry profiles into a workflow engine
available at the beamline which in turn can communicate directly with an
industrial AI inference server (here \emph{Nvidia Triton} deployed on
VISA hardware).

We emphasize again that for both described schemes no additional software
needs to be installed into the beamline control software environment.
The first, TANGO-based, approach largely benefits from the independence
regarding the specific beamline control software solution and therefore its
transferability to other facilities with a TANGO support layer on
beamlines (e.g. DESY, SOLEIL or ALBA). Using industrial AI
inference server in the second approach instead targets the use of
standardized API interfaces commonly used in ML/AI.

\subsection{Machine learning methods}

While the focus of the present paper is not on the specific details of
ML methods for X-ray data analysis, we briefly outline the concepts
employed. For further details we refer to \cite{greco_fast_2019, greco_neural_2021, greco_neural_2022, marecek_faster_2022, munteanu_reflectorch_2023 }.
For fast automated analysis of the measured XRR and Bragg reflection
data, we rely on neural network-based maximum likelihood estimation
(MLE). Compared to previous implementations, here we incorporate prior
knowledge about the sample at a given time into the machine learning
pipeline, thereby effectively mitigating uncertainty and constricting
the range of potential solutions. For a comprehensive, detailed discussion of this approach, see \cite{munteanu_reflectorch_2023}.

The reflectometry analysis aims at reconstructing the scattering length
density (SLD) profile of the studied sample based on the measured
reflectivity curve. Given an SLD profile, the corresponding theoretical
curve can be swiftly calculated \cite{parratt_surface_1954}. However, reversing this
operation presents a challenge because of the inherent ambiguity that
often allows for multiple, different SLD profiles to correspond to the
same curve within the bounds of measurement uncertainty. Fundamentally, this is related to the famous phase problem of scattering (since only the intensity, but not the phase is recorded in the detector).  Consequently,
it is vital in reflectivity analysis to make use of the physical understanding of
the investigated system in order to reduce the number of potential
solutions and identify the correct one. In previous ML-based works with
two-layered structures \cite{greco_fast_2019, hinderhofer_machine_2023}, we approached this task by effectively fixing most of the
parameters characterizing the SLD profile and training the neural
network to estimate only the three unfixed parameters - thickness,
roughness, and density of the top organic layer - anticipated to vary
among the samples under study, with parameters of the silicon substrate
with a silica top layer held constant. Expanding this method to
accommodate a larger set of variable parameters, further techniques to
address the ambiguity problem are needed \cite{munteanu_reflectorch_2023}. In the present study, we build
upon this approach and showcase two potential methods for integrating
physical knowledge into the machine learning framework.
    
First, we propose to include the boundaries of the open parameters as an
additional input to the neural network. As before, for each open
parameter $\theta_{i}$ we designate broad ranges
$[\theta_{i}^{\textit{total min}},\theta_{i}^{\textit{total max}}]$
that determine the general scope of the neural network. Yet, in
conjunction with these fixed ranges, we introduce
\emph{sample-dependent} ranges
$[\theta_{i}^{\min}, \theta_{i}^{\max}]$ 
$(\theta_{i}^{\min} \leq \theta_{i}^{\max}; \theta_{i}^{\min}, \theta_{i}^{\max} \in [\theta_{i}^{total\ min}, \theta_{i}^{total\ max}])$
that impose constraints on the fitted parameters for each sample studied
and are supplied as additional input to the neural network alongside the
measured curve. This method effectively confines the solution space for
a particular sample while maintaining extensive overall parameter ranges
within a single neural network. In instances where a single solution
exists within the provided sample-dependent boundaries, the inverse
operation becomes well-defined, enabling a precise fit. We note that
this approach effectively combines the best of the two worlds: the
flexibility of the parameter ranges of the conventional differential
evolution method, and the speed of the neural network. Furthermore, for
the real-time \emph{in situ} analysis, it allows harvesting the
information from the previous fits, as prior knowledge, so that the analysis of curves
measured at different times for the same sample are no longer
independent. For instance, during the growth experiment, we define the
sample-dependent constraints on the growing film thickness based on the
obtained result from a previous fit. This way, we leverage the physical
understanding of the growth process (thickness of the growing layer
cannot decrease over time) and the experimental setup (thickness cannot
increase too rapidly), without a specific growth model restriction as
used in \cite{marecek_faster_2022}.

\begin{figure}
\caption{Parameterization of the periodic multilayer structure. The stacked layers share the same thickness $d_{\mathrm{monolayer}}$, roughnesses, and SLD densities. Most of the parameters introduced are relative to $d_{\mathrm{monolayer}}$. Two sigmoid functions modulate the resulting periodic SLD profile. The position of the first sigmoid $c_{\mathrm{sig}}$ defines the number of monolayers and the corresponding thickness $d_{\mathrm{film}} = d_{\mathrm{monolayer}} \cdot c_{\mathrm{sig}}$, while its width $\sigma_{\mathrm{sig}}$ determines the surface roughness $\sigma_{\mathrm{film}} = \sigma_{\mathrm{sig}} \cdot d_{\mathrm{monolayer}}$. The second sigmoid modulates the contrast $\Delta\rho$ between the layers, resulting in smoother SLD profile at the surface.}
\includegraphics[width=\textwidth,keepaspectratio]{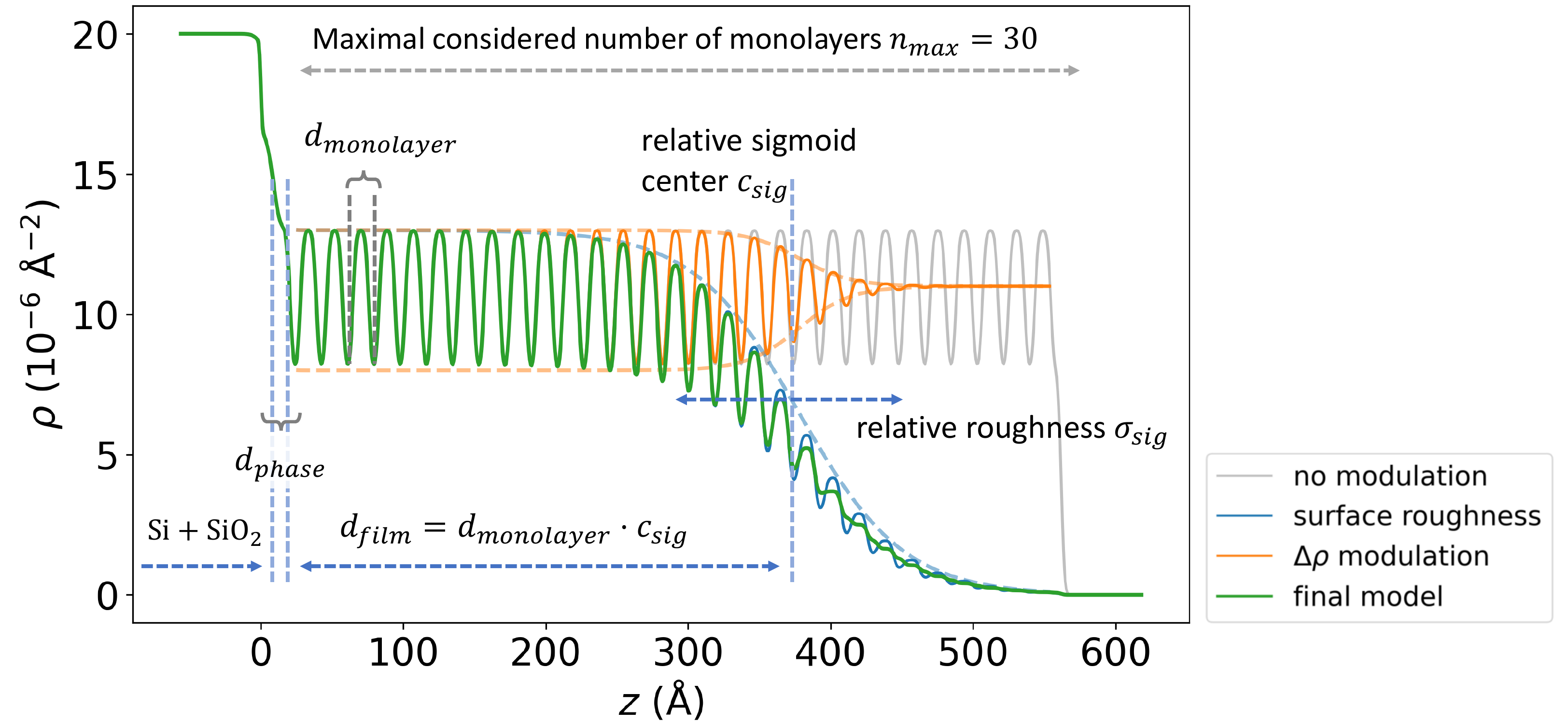}
\label{fig:sld}
\end{figure}

The second method for integrating physical knowledge into the ML
framework that we employ in this work is physics-based parameterization.
In the case of ML-based reflectivity analysis, this approach was first
introduced by \cite{marecek_faster_2022}, where the physics-based
growth model allows to effectively reduce the number of estimated
parameters. In this work, we apply this approach to the case of periodic
multilayer structures by implementing a physics-based parametrization of
the SLD profile. The standard parameterization of the box model implies
three parameters (or four parameters when including absorption) per
single layer for each box (density, thickness, roughness). Given that a
single molecular monolayer block is typically modeled by two layers,
such parameterization would require up to
$2 \cdot \ 30\  \cdot \ 3 = 180$ independently fitted parameters (when considering 30 monolayers)
resulting in a $\mathbb{R}^{180}$ solution space. However, such
parameters are largely correlated, because the monolayers consist of the
same material, feature the same thickness, etc. To provide this information to the neural
network, we introduce a set of $17$ independently fitted or predicted
parameters, that together define all the 180 parameters of the box
model. Such parameterization is based on a physical understanding of
monolayer structures and significantly decreases the task's complexity \cite{marecek_faster_2022}. The parametrization is
illustrated in Fig. \ref{fig:sld} and is discussed in more detail in \cite{munteanu_reflectorch_2023}.

We have developed a fast, GPU-accelerated module within the PyTorch framework that calculates reflectivity curves in order to train the neural networks utilized in the experiment in real-time during the beamtime. This module seamlessly integrates into the
training process, enabling on-the-fly XRR simulations throughout the
training phase. Consequently, we were able to quickly adjust the training settings (even during the beamtime), eliminating the requirement for
meticulous pre-planning. Note, that we rely on preprocessing operations
and the rich postprocessing pipeline introduced in \cite{greco_neural_2021, greco_neural_2022} that features $q$ offset sampling and LMS fit. A major
addition to the postprocessing step of the previous works is the
introduction of a new optional fitted parameter that models a linear
change of thickness \emph{during the measurement of a single curve} as
we perform \emph{in situ} and real-time experiments. Therefore, for a single scan, we
take into account the time-dependent layer thickness $d(t)$ and the
time-dependent incidence angle corresponding to momentum transfer
$q(t)$. Together with other time-independent parameters
$\theta$, $d(t)$ and $q(t)$ define the curve:
$R =  R ( q ( t ), d ( t ), \theta )$.
A linear model of a growth process
$d( t ) = d_{0} +  t\frac{\Delta d}{\Delta t}$ is a very
good approximation for the short duration of a quick real-time scan.
Consequently, we fit two parameters $d_{0}$ and
$\frac{\Delta d}{\Delta t}$ for each curve via LMS, using the neural
network prediction as an initial guess for the parameter
$d_{0}$. The introduction of this time-dependent parameter is
required to accurately fit XRR curves measured during the growth in the
case of a fast growth process relative to the data acquisition time for a
single XRR profile, as Kiessig fringes will be slightly narrower at high
$q$ as compared to low $q$.

\subsection{Data handling}

All data, including the extracted parameters of the online data
analysis, have been stored in a fully correlated fashion, i.e. raw data
alongside with analysis results in NeXus HDF5, and are publicly
available through the ESRF data portal \cite{pithan_controlled_2023}.

Using the ESRF file-saving infrastructure (Bliss + NeXusWriter)
enabled the visiting experimentalists to directly insert the online data
analysis results into the data files published by the beamline (see
supporting information for details). Furthermore, through integrating into
the ESRF beamline software ecosystem, the online data analysis results could
seamlessly be transferred into the accompanying electronic logbook, allowing a first overview of the experimental results ordered via timestamps.


\subsection{Experiment}

This study has been performed on the surface scattering branch of the
ESRF ID10 beamline \cite{jankowski_complex_2023}. A beam energy of
$17.0 keV$ and beamsize of $30x30 \mu m$ was used. We use X-ray reflectometry (XRR), an established surface scattering techniques, performed following standard procedures \cite{daillant_x-ray_2009,tolan_x-ray_1999,holy_high-resolution_1999,seeck_observation_2002}. A user supplied UHV sample
environment (Fig. \ref{fig:dataflow}) has been installed in horizontal geometry.
Molecular thin film samples are prepared \emph{in situ} using molecular
beam deposition \cite{ritley_portable_2001, zykov_diffusion_2017}.

To achieve the objective of stabilizing and terminating the
self-assembly, growth, and crystallization characteristics of molecular
thin film studies \emph{in situ}, a closed-loop control system has been
implemented, leveraging the ML-based online analysis. This closed-loop
control allows for autonomous experiments. In this particular study, the
ML-based closed loop system assumes control over the operation of two
shutters, which involves covering either the substrate or the incoming
molecular beam (refer to Fig. \ref{fig:dataflow}). To limit beam damage, which might
occur for longer scan times at lower deposition rates, we reduced the
X-ray flux to a level where there was no noticeable impact on Bragg peak
intensities over the time of a growth run at our deposition rates.

In this study, molecular thin films of aluminium-tris(8-hydroxychinolin)
(\Alq, $C_{27}H_{18}AlN_{3}O_{3}$),
a frequently used component in organic light-emitting diodes, were grown to serve
as an exemplary material system for amorphous molecular thin films \cite{ mondal_effect_2021}. To demonstrate the online capabilities regarding
the analysis of crystalline multilayer systems the organic semiconductor
N,N`-Dioctyl-3,4,9,10- perylendicarboximid (\PTCDIC ,
$C_{40}H_{42}N_{2}O_{4}$)
\cite{zykov_diffusion_2017, krauss_three-dimensional_2008} was chosen for
demonstration purposes. For details on the scientific background of
these materials, we refer to \cite{kowarik_organic_2008,schreiber_organic_2004,witte_growth_2004}.

\section{Results and discussion}

To verify and test the technical solutions discussed above, we
aim to grow molecular thin films of predefined thickness where the
ML-based closed loop takes control over the termination of the growth
process by closing the relevant deposition shutters. It became evident that
it is crucial to provide prior knowledge about the film parameters (e.g. plausible
film thickness ranges) as input of the ML-model to achieve robust
fitting for a large number of consecutive scans. In this work priors were
used to describe minimum and maximum boundaries for each parameter.

\subsection{Discussion of ML in XRR and Bragg reflection fitting}

To achieve good performance of ML predictions on XRR signals it
is important to consider corrections such as $q$-shift
(for slight misalignment) and subsequent fitting using a LMS algorithm as described in \cite{greco_neural_2021}. Furthermore, due to the
\emph{in situ} data-taking that is inherent to closed loop feedback, the
varying film thickness during a single scan must be taken into account
if the scan speed is not fast compared to the deposition speed. We
accommodate this additional effect as an additional parameter in the
subsequent LMS fit as discussed above (snippet in Supporting Information section SI-4). In both
examples of amorphous and multilayer thin films, prior information was
taken into account for the online ML analysis result: In the multilayer fits
physical knowledge was embedded in the parametrization of the electron
density model and rather narrow training ranges of the ML model. In
contrast, wide training ranges were used for the single-layer model, but
prior information was taken into account by inputting it into the ML model
to achieve a certain regularization of the results. Here, in the dataset
presented in Fig. \ref{fig:singlelayer}, through the use of priors, the expected film
thickness was constrained to increase by at most 50 Å and decrease by no
more than $25$ Å for the ML based real-time analysis. This still gives the model
sufficient flexibility to predict changes during the film growth, but
imposes weak boundaries based on the scan speed and deposition rate.

\subsection{Single layer fits (Kiessig oscillations)}

\begin{figure}
\caption{ML controlled thin film deposition: a) Exemplary measurements
and fits. The experimental data are acquired using fast, real-time scans,
thus the poor counting statistics around $q\simeq 0.1 1/$Å is due to the
limited number of absorber changes. b) target thicknesses in closed loop
operation vs. actually measured thicknesses after the closed loop
feedback terminated the growth.}
\includegraphics[width=.45\textwidth]{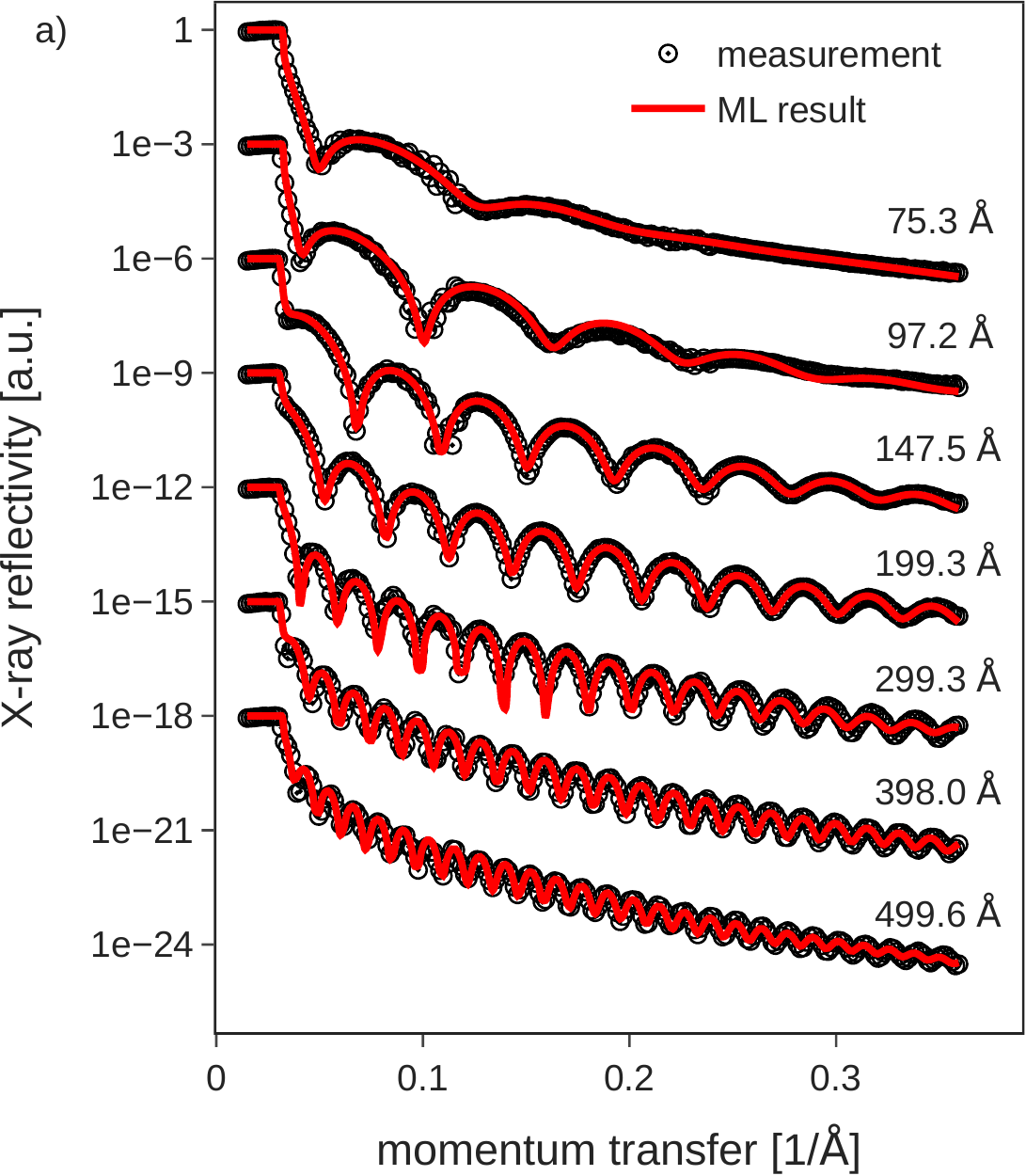} \\
\includegraphics[width=.45\textwidth]{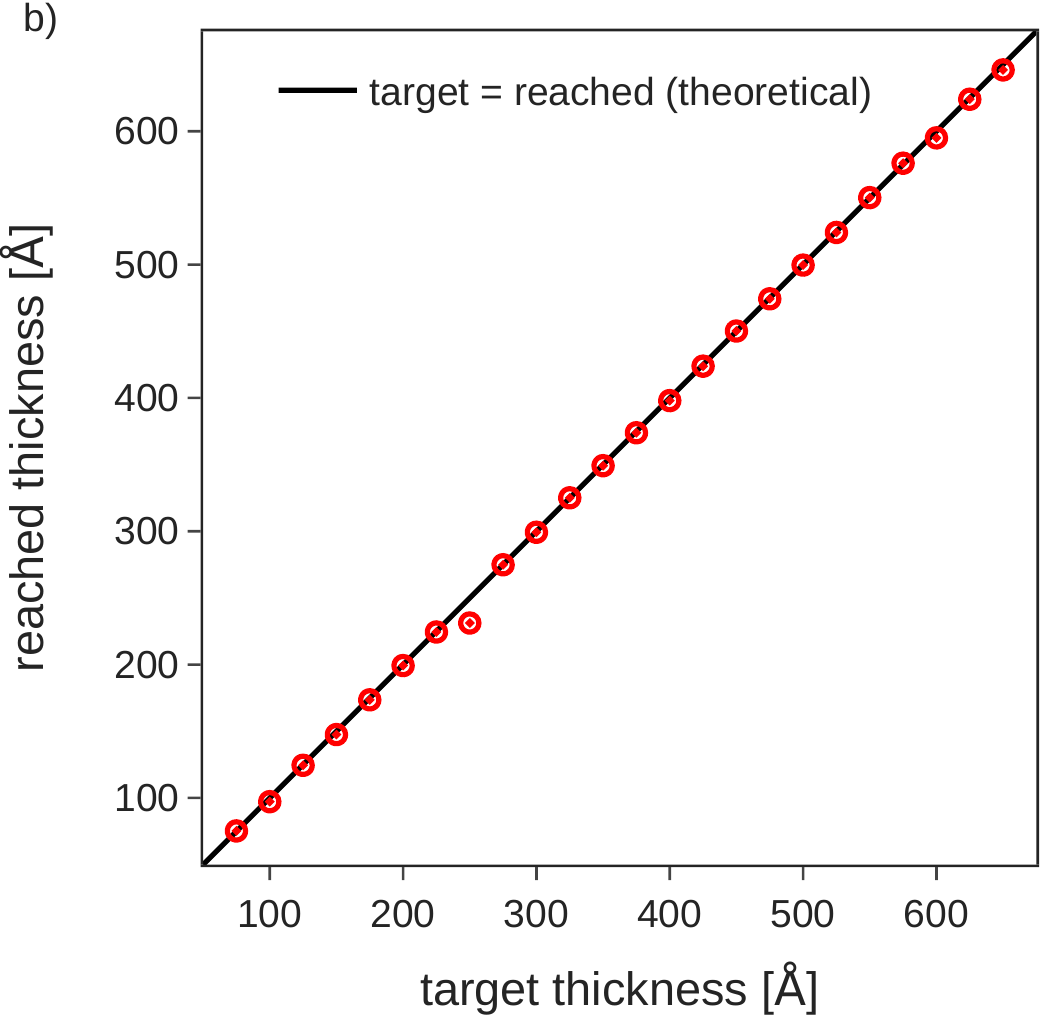}
\label{fig:singlelayer}
\end{figure}

In Fig. \ref{fig:singlelayer}a we show exemplary reflectometry profiles acquired during thin
film growth of \Alq using continuous scans (also known as
\emph{flyscans} or \emph{fastscans}) together with the corresponding ML results of
layer parameters. As can be seen from the presented, footprint corrected
scans, we achieved a very good fit quality using the ML-approach with
priors. Both the period of Kiessig fringes as well as the roughness
induced damping of Kiessig fringes are correctly reproduced in the ML
fits. This is a prerequisite for achieving closed loop control to
terminate growth at the target thickness, but due to the finite duration
of the scans ($45s$ per XRR profile in Fig. \ref{fig:singlelayer}a) and an average growth rate
of $1$ \textsf{nm/min}, the target thickness may be reached during a scan.
Therefore, we used a linear extrapolation of the thickness information of
previous scans to automatically calculate the best moment to close the
shutter, as yet another asynchronous process (see Supporting Information section SI-3). Fig. \ref{fig:singlelayer}b shows the result of the closed loop deposition control for several
target thicknesses between $80$ Å and $640$ Å. The target thicknesses are
plotted on the x-axis, while the true film thicknesses at which the
deposition was terminated are given on the y-axis. As expected for a
functioning closed loop control, the data indeed falls on the diagonal
line, except for one outlier. Overall, the chosen target
thicknesses could be reached within $\pm 2$Å ($0.1 \% $) average accuracy.

\subsection{Multilayer fits (Bragg reflections)}

\begin{figure}
\caption{a) ML-based fits of \PTCDIC multilayer structures. As input
for the ML-model only points measured around the \PTCDIC Bragg-peak were
taken into account to enable faster data acquisition. b) scattering
length density profiles corresponding to a).}
\includegraphics[width=.42\textwidth]{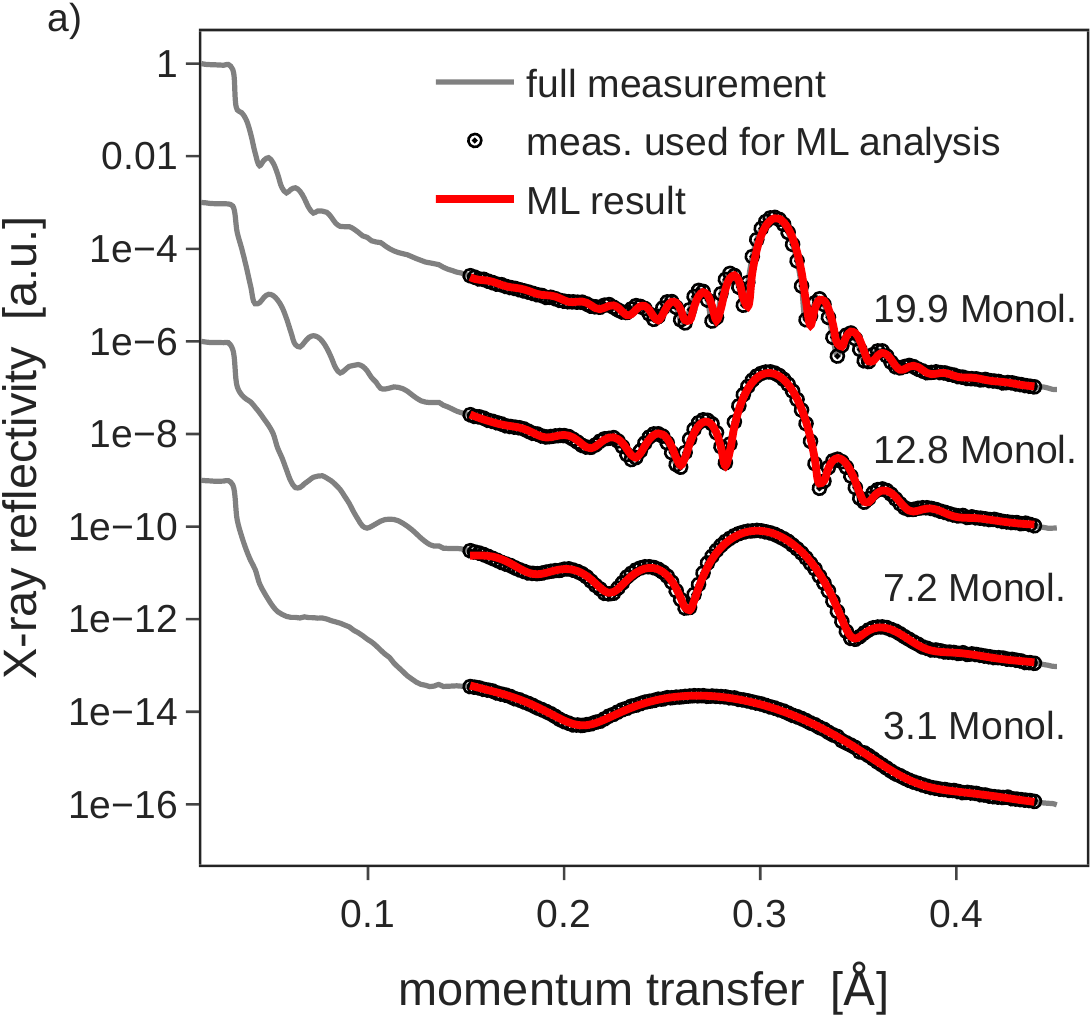}\qquad
\includegraphics[width=.42\textwidth]{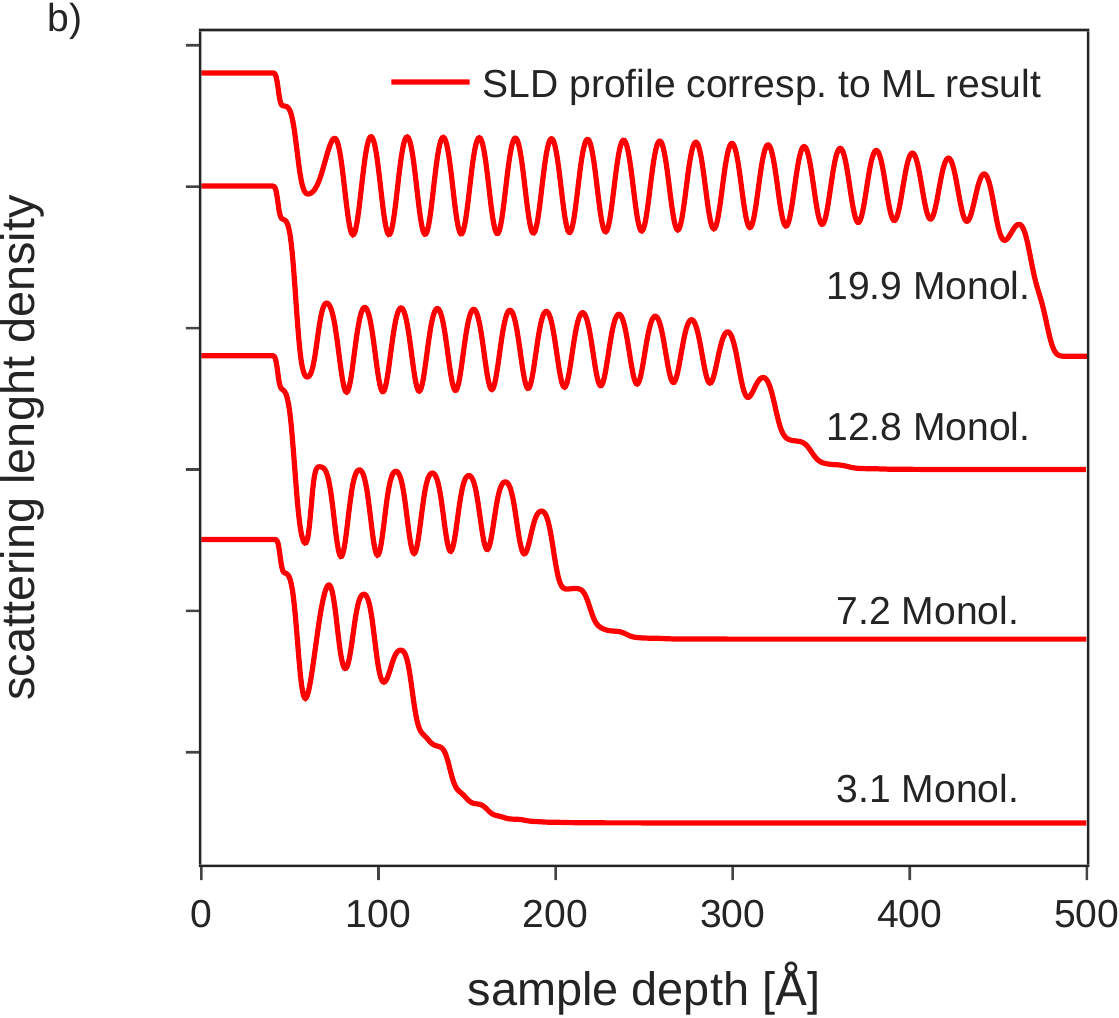}
\label{fig:multilayer1}
\end{figure}

Not only amorphous thin film structures, but also multilayers of the
molecule \PTCDIC were studied. By incorporating
physical knowledge of the sample structure into the parametrization of
the used ML model we are able to fit Laue fringes and the molecular
Bragg-peak that arises with increasing film thickness from molecular
multilayers (Fig. \ref{fig:multilayer1}a). To speed up data acquisition only a relatively
short $q$-range centered around the molecular Bragg peak was
repeatedly scanned while running in closed loop mode. An initial 'full'
XRR curve, including the total reflection edge, was used for signal normalization before activating the closed loop operating mode.

\begin{figure}
\caption{a) Thin film thickness extracted through ML model in real
time during the measurement compared to a quartz crystal micro balance (QCM)
reference. For films with thickness above $\simeq 250$ Å, i.e. 
$\simeq$ 12 mol. monolayers ML-results are reliable. b) selected fits from the real-time dataset. Ambiguity in the thickness
predictions for lower thicknesses mainly results from the absence of
significant features in the investigated q-range.}
\includegraphics[width=.42\textwidth]{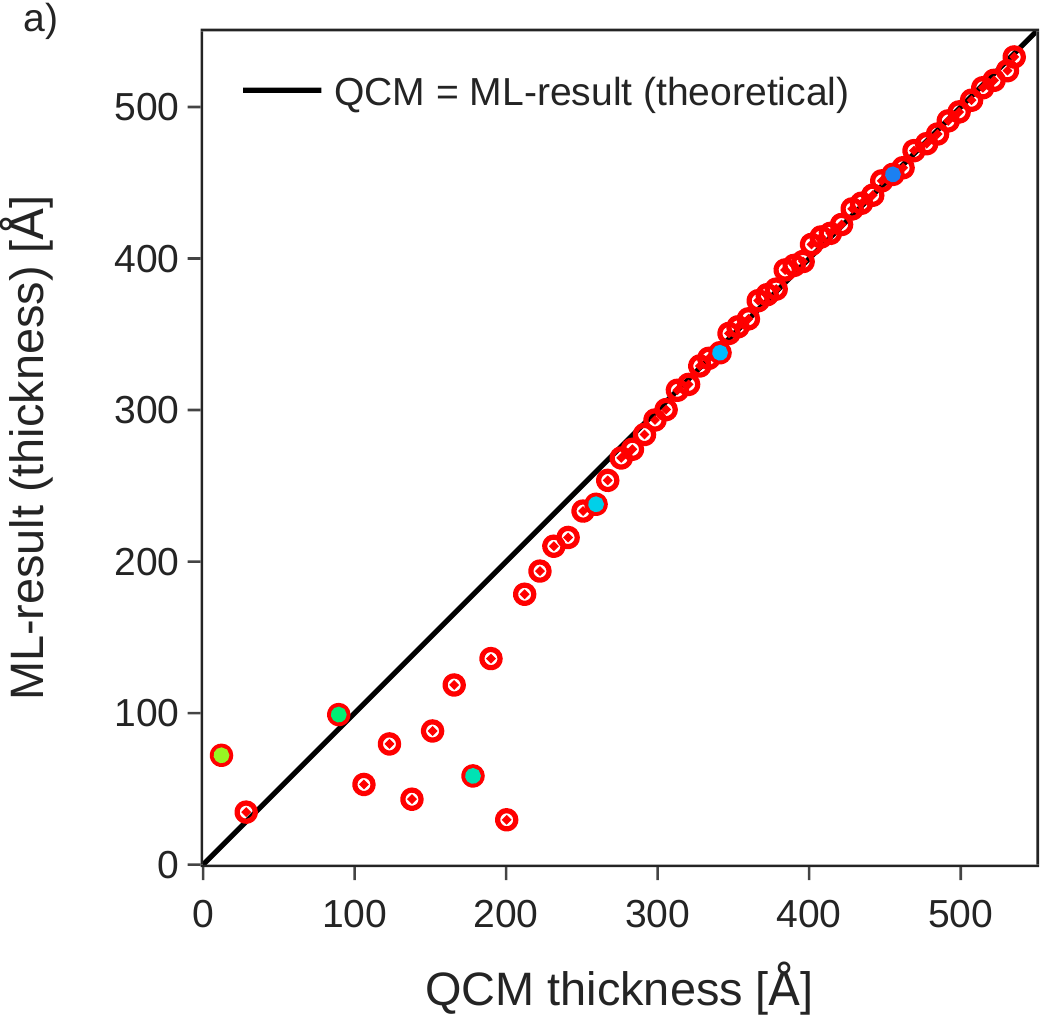} \qquad
\includegraphics[width=.44\textwidth]{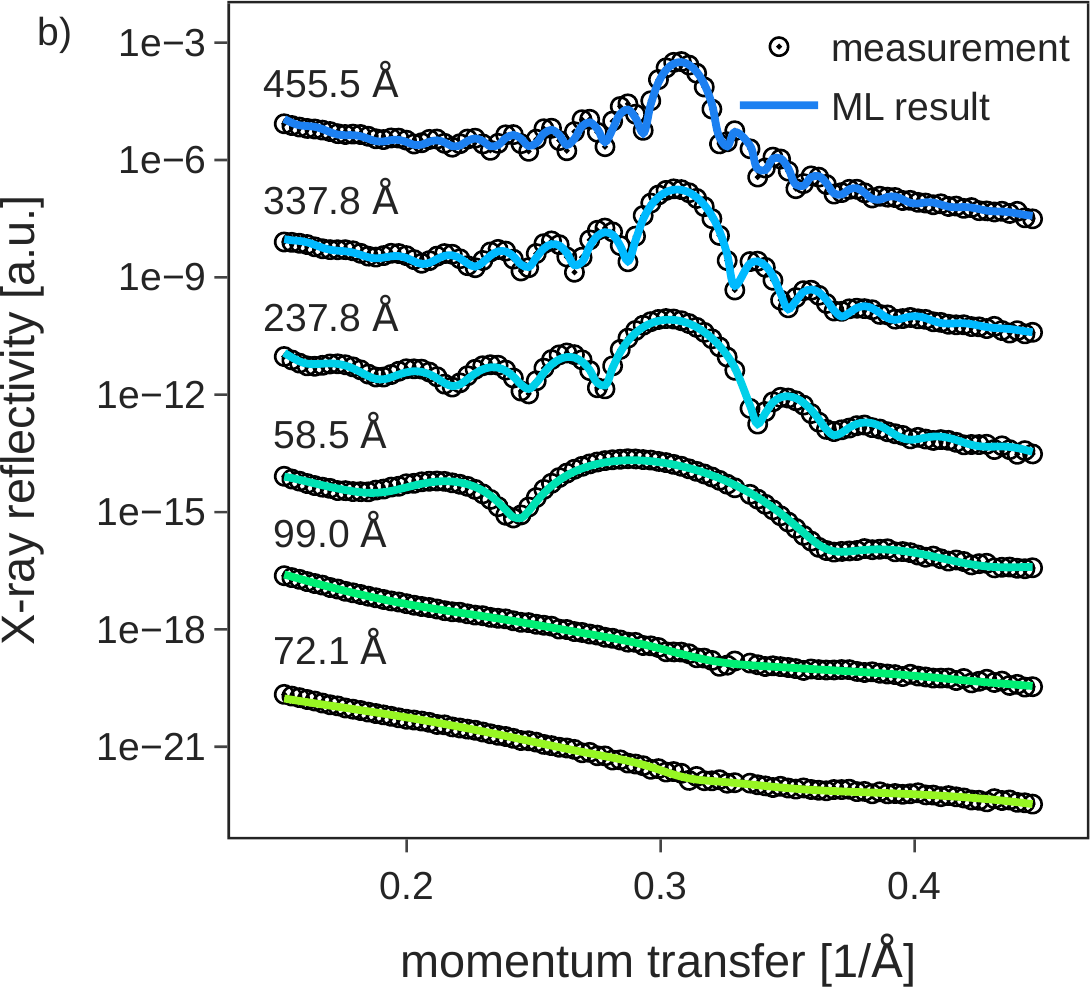}
\label{fig:multilayer2}
\end{figure}

Again, these scans around the Bragg reflection could be fitted by ML with
high fidelity, reproducing the Bragg peak and the period of the Laue
oscillations and their asymmetric intensity to the left and right of a
Bragg reflection. The corresponding electron density profiles from the
live fits are shown in Fig. \ref{fig:multilayer1}b, from which one can directly infer the
number of deposited monolayers, as one oscillation of the scattering
length density corresponds to one \PTCDIC monolayer.
Comparing the film thickness from the ML fits of the Bragg region to the
total thickness derived from a measurement of the deposition flux with a
quartz crystal microbalance (QCM) one again finds reasonable agreement
as shown in Fig. \ref{fig:multilayer2}a). For thicknesses above 250 Å the agreement is
good. Some scatter is visible below this thickness, even though
the quality ML fit of the individual reflectometry curves is good (Fig. \ref{fig:multilayer2}b) ). This particular example shows the difficulty
of ambiguous XRR fits, where several sample structure models
can fit a single X-ray curve. Also, note, that the QCM measures total
film thickness only if a constant sticking factor on the substrate and previously deposited molecular material is assumed. Further, the Laue oscillations do not correspond to the total
film thickness, but the coherently ordered film thickness, so initial
disorder in the crystal lattice may to some degree contribute to the
observed deviations for low film thicknesses (additionally, of course, the potentially non-integer layer occupancy during growth may interplay as well \cite{rieutord_x-ray_1989}). Overall, we conclude that
the ML results (live) of the coherent film thickness during data acquisition
nicely match our detailed post growth analysis and yield consistent data
for larger film thickness of our sample system. This, in principle,
enables closed loop feedback e.g. to target growth of a fixed number of
crystalline lattice planes in a thin film sample.

\subsection{Robust feedback to control growth}

Within the domain of ML-based closed-loop feedback, we emphasize the
criticality of robust real-time analysis for successful operation. Especially in
Fig. \ref{fig:singlelayer} the expected trend of increasing thickness from scan to scan
during growth is clearly observed. The robustness of the ML results is
especially important when facing the challenge on the control side that
extracted parameters must be used to extrapolate the film thickness into
the near future. Only robust fits can lead to a meaningful time-series,
so that after extrapolation the autonomous growth termination can act at
the moment the predefined film thickness is reached -- which does not
necessarily coincide with the end of a performed scan. In this study,
the time resolution of the reflectivity measurements (repeat rate of
scans) could be identified as the most critical bottleneck of the given
setup. Therefore, the limited time resolution had to be addressed
through an additional asynchronous process taking care of the temporal
extrapolation and the triggering of actions in close loop operation.
Here BLISS is very well equipped for this kind of tasks through its
tight integration of \emph{gevent} to enable coroutine operation \cite{bilenko_gevent}. Through averaging 
several of the last ML-results (a.k.a. ``predictions''), isolated slight outliers in the ML results could be tolerated and were not sufficient to invalidate the live feedback
mechanism. Overall, the $2\%$ thickness error proves the robust ML
fitting, extrapolation and closed loop feedback action.

\subsection{Further integration potential}

For the experiment conducted in the context of this work it was possible
to store ML analysis results together with the original raw data and
to interact with the facility-provided electronic logbook (ELN). However, ingesting machine-readable metadata into the facility
data catalog remained difficult. In order to support and contribute to a
thriving ecosystem of ML models for X-ray data analysis, i.e. physics-informed ML models and sample-system specific ML models, it is of crucial
importance that facilities provide -ideally standardized- interfaces and
best practice guidelines on how externally developed code should
interact with the respective beamline control and data storage systems.
Initiatives similar to DAPHNE4NFDI \cite{daphne4nfdi} are well suited for discussion in this context, since they bring together facilities and the user community,
including experienced users, to jointly design research data infrastructure
along a process chain all the way from the proposal and experiment to
the fully analyzed and archived data.

\section{Conclusions and outlook}

In this study we established a complete closed loop feedback cycle for
controlling thin film and crystal growth exclusively relying on
real-time scattering data and online ML analysis. The presented scheme
is well suited for a broader range of \emph{in situ} and \emph{in
operando} experiments. Not only growth dynamics can be observed with
X-rays, but also processes in dynamic equilibrium where the information 
extracted from the scattering data itself can be used to stabilize the
equilibrium. Specifically, for XRR this may for instance involve control over various liquid systems with a Langmuir trough as one possibility to prepare environments (e.g. lipid bilayers) with precise surface pressure to have stable conditions at the liquid-air interface. Here the ML approach for feedback directly relates to the film properties measured with XRR, whereas of course the sample environment (e.g. surface pressure of a Langmuir trough) may also be controlled externally.
In a broader picture, we see a bright future to embed ML based feedback loops
also for other types of scattering experiments e.g. involving
electrochemical control over battery charging or
control over electrochemical sample environment conditions or catalytic
conditions in experiments on nanoparticles. It also extends to other synchrotron-based techniques \cite{chen_machine_2021, hinderhofer_machine_2023},
as well as neutron-based techniques, including in particular
neutron reflectometry (NR) \cite{treece_optimization_2019}.

With the focus on publicly available FAIR datasets hosted in facility
data catalogs we see the potential of ML based online data analysis
in helping to make these datasets in catalogs ``more fair and more reusable''.
This is possible through enriching the currently archived raw dataset at least with the
preliminary analysis results on the fly. Live ML X-ray
data analysis has the ability to complete datasets with scientifically
relevant machine-readable metadata as well as automated capture of
scientific results in electronic logbooks (ELNs) accompanying the
dataset in the facility data catalog. We are convinced that the
presented approach can contribute significantly to a more efficient use
of beamtime at large scale facilities. We envision that integration of
live data analysis and feedback loops with ML models will become more
established at beamlines along the lines presented here. Then, facility
users can use ML to observe live how experiments progress and also perform
previously unattainable experiments with direct feedback and contribute
to an ever-growing, meaningful X-ray dataset pool. 

\appendix

\section{Data Availability Statement}

The specific beamline integration python scripts as well as the TANGO server 	embedding the ML model used in this study are available in the Supporting Information accompanying the published article. Data underlying this publication is available in the ESRF data repository \cite{pithan_controlled_2023}. Experimental data used to prepare for the beamtime is available in \cite{pithan_2022}. A reference ML model is available in \cite{greco_2022_zenodo} and soon to be found in \cite{munteanu_reflectorch_2023}.

\section{\ack{Acknowledgements}}

We acknowledge the European Synchrotron Radiation Facility (ESRF) for providing of synchrotron radiation at the ID10 beamline. Further, we express our gratitude to those offering support with the ESRF computing infrastructure, especially Jean-Francois Perrin, Wout de Nolf, Samuel Debionne and Valentin  Valls.\\
We also acknowledge financial support via the DAPHNE4NFDI project and the BMBF. 

\section{Glossary}

\begin{center}
\begin{tabular}{ |l|r| } 
 \hline
BLISS & ESRF experiment control system https://gitlab.esrf.fr/bliss/bliss \\
LIMA & LImA : Library for Image Acquisition (ESRF) https://lima1.readthedocs.io \\
SCADA & supervisory control and data acquisition system \\
TANGO & SCADA system originally developed at the ESRF https://www.tango-controls.org \\
 \hline
\end{tabular}
\end{center}

\bibliographystyle{iucr}
\bibliography{closed_loop}



\section*{Supplementary material}

At the time of the experiment BLISS 1.10 and pytango 9.3 were available on the ID10 beamline.

## SI-1. Tango device servers wrapping around ML Models

A tango device server with specific API to wrap around the ML models developed based on pytorch and tensorflow used in the context of this study.

```python
from typing import Iterable
import numpy as np
from pprint import pprint

from tango.server import Device
from tango.server import run
from tango.server import attribute, command, device_property
from tango import AttrWriteType

# python code wrapping around pytorch model
# available on request or in follow-up publication
from reflectorch.inference import InferenceModel


def energy_to_wavelength(energy: float):
    """Conversion from photon energy (eV) to photon wavelength (
    angstroms)"""
    return 1.2398 / energy * 1e4


class TorchEvaluator(Device):
    preprocess_attributes = {
        "beam_width": np.float64,
        "sample_length": np.float64,
        "energy_keV": np.float64,
    }

    input_attributes = {
        "tth": [np.float64],
        "intensity": [np.float64],
        "transm": [np.float64],
        "priors": [[np.float64]],
        "count_time": [np.float64],
    }
    prediction_attributes = {
        "q": [np.float64],
        "q_interp": [np.float64],
        "refl": [np.float64],
        "refl_interp": [np.float64],
        "refl_predicted": [np.float64],
        "refl_predicted_polished": [np.float64],
        "parameters": [np.float64],
        "parameters_polished": [np.float64],
        "parameter_names": [str],
```

```python
            "sld_profile": [np.float64],
            "sld_x_axis": [np.float64],
            "sld_profile_polished": [np.float64],
    }

    def __init__(self, *args, **kwargs):
        Device.__init__(self, *args, **kwargs)

        self._prefix_dict = dict()

        self._input_data = self._initialize_dynamic_attributes(
            "input", self.input_attributes
        )
        self._prediction_output = self._initialize_dynamic_attributes(
            "prediction", self.prediction_attributes
        )
        self._preprocess_data = self._initialize_dynamic_attributes(
            "preprocess", self.preprocess_attributes
        )
        self.inference_model = InferenceModel("l2q256_exp_3")

    @property
    def prediction_output(self):
        return self._prediction_output

    @command(dtype_out=int)
    def predict(self) -> dict:
        """
        blocking command that runs pytorch model inference
        """
        intensity = self._input_data["intensity"]
        scattering_angle = self._input_data["tth"]
        attenuation = self._input_data["transm"]
        priors = self._input_data["priors"]

        preprocess_params = {
            "wavelength": energy_to_wavelength(
                self._preprocess_data["energy_keV"] * 1000
            ),
            "beam_width": self._preprocess_data["beam_width"],
            "sample_length": self._preprocess_data["sample_length"],
        }

        self.inference_model.set_preprocessing_parameters(**
preprocess_params)

        res_dict = self.inference_model.predict(
            intensity, scattering_angle, attenuation, priors,
        )

        pprint(res_dict)

        self.prediction_output["q"] = res_dict["q_values"]
        self.prediction_output["q_interp"] = res_dict["q_interp"]
        self.prediction_output["refl"] = res_dict["curve"]
        self.prediction_output["refl_interp"] = res_dict["curve_interp"]
```

```
        ]
99      self.prediction_output["refl_predicted"] = res_dict["
    curve_predicted"]
100     self.prediction_output["parameters"] = res_dict["params"]
101     self.prediction_output["sld_profile"] = res_dict["sld_profile"]
102     self.prediction_output["sld_x_axis"] = res_dict["sld_x_axis"]
103     self.prediction_output["parameter_names"] = res_dict["
    param_names"]
104
105     self.prediction_output["refl_predicted_polished"] = res_dict["
    curve_polished"]
106     self.prediction_output["parameters_polished"] = res_dict["
    params_polished"]
107     self.prediction_output["sld_profile_polished"] = res_dict[
108         "sld_profile_polished"
109     ]
110
111     return 0
112
113 def _initialize_dynamic_attributes(self, prefix, attributes):
114     """
115     boiler plate code to achieve generic handling of data arrays
    and
116     parameters that have to be passed into the tango server and
    back
117     to the client
118     """
119
120     if prefix in self._prefix_dict:
121         att_dict = self._prefix_dict[prefix]
122     else:
123         att_dict = dict()
124
125     for k, v in attributes.items():
126         att_type = None
127
128         if isinstance(v, tuple):
129             att_type = v[0]
130             att_dict[k] = v[1]
131         elif isinstance(v, Iterable):
132             att_type = v
133             att_dict[k] = []
134         else:
135             att_type = v
136             att_dict[k] = None
137
138         attr = attribute(
139             name=prefix + "_" + k,
140             dtype=att_type,
141             access=AttrWriteType.READ_WRITE,
142             fget=self._generic_read,
143             fset=self._generic_write,
144             max_dim_x=256,  # self.N_MAX_X,
145             max_dim_y=256,  # self.N_MAX_Y,
146             # fisallowed=self.generic_is_allowed,
147         )
```

```
148                self.add_attribute(attr)
149
150        self._prefix_dict[prefix] = att_dict
151        return self._prefix_dict[prefix]
152
153    def _generic_read(self, attr):
154        prefix, name = attr.get_name().split("_", 1)
155        value = self._prefix_dict[prefix][name]
156        # unlike a normal static attribute read, one has to modify the
        value
157        # inside this attr object, rather than just returning the value
158        attr.set_value(value)
159
160    def _generic_write(self, attr):
161        prefix, name = attr.get_name().split("_", 1)
162        self._prefix_dict[prefix][name] = attr.get_write_value()
163
164
165 if __name__ == "__main__":
166    run((TorchEvaluator,))
```

**SI-2. Beamline / BLISS integration to run ML inference on remote tango device server and publish prediction results together with raw data**

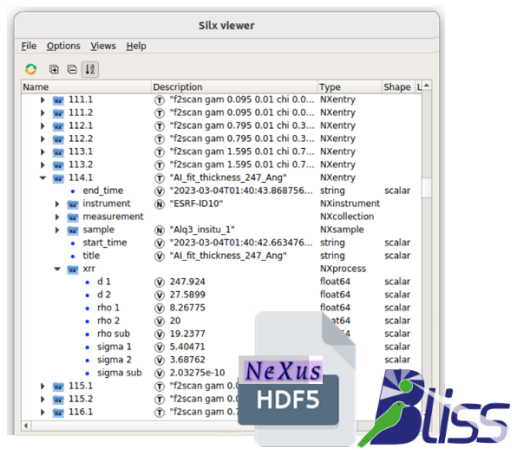

Fig. SI-1.: NeXus compliant data file generated by the beamline control system including results of the online data analysis.

Script that can be loaded as BLISS user macro into the beamline software environment without any additional external dependencies. The function predict_and_save takes a bliss scan object as input, runs ML inference and processes the results so that they are saved together with the raw data in a coherent fashion.

References:

https://bliss.gitlab-pages.esrf.fr/bliss/1.10.x/scan_group.html#sequences (accessed: 05.2023)

https://bliss.gitlab-pages.esrf.fr/bliss/1.10.x/data/data_scan_metadata. html (accessed: 05.2023)

```python
import numpy as np
from dataclasses import dataclass

from bliss.common import tango
from bliss import setup_globals
from bliss.scanning.group import Sequence
from bliss.scanning.chain import AcquisitionChannel
from bliss.scanning.scan_meta import USER_SCAN_META
# Add XRR Metadata entry to BLISS
if "xrr" not in USER_SCAN_META.categories_names():
    USER_SCAN_META.add_categories({"xrr"})
    USER_SCAN_META.xrr.set("xrr", {"@NX_class": "NXprocess"})


@dataclass
class Priors:
    d_top: tuple = (0.0, 1000.0)
    d_sio2: tuple = (10.0, 30.0)

    sigma_top: tuple = (0.0, 50.0)
    sigma_sio2: tuple = (
        0.0,
        10.0,
    )
    sigma_si: tuple = (
        0.0,
        2.0,
    )

    rho_top: tuple = (7.0, 14.0)
    rho_sio2: tuple = (19, 23)
    rho_si: tuple = (19, 21)
    max_d_change: float = 5.0
    fit_growth: bool = True

    def to_arr(self):
        return np.array(
            [
                self.d_top,
                self.d_sio2,
                self.sigma_top,
                self.sigma_sio2,
                self.sigma_si,
```

```
45                    self.rho_top,
46                    self.rho_sio2,
47                    self.rho_si,
48                ]
49            )
50
51
52  @dataclass
53  class PreprocessParams:
54      energy_keV: float = 17.0
55      beam_width: float = 0.03
56      sample_length: float = 9.0
57
58
59  def predict_and_save(
60      scans, priors=None, preprocess: PreprocessParams = None,
61  ):
62      """
63      takes a list of scans that will be concatenated to a single XRR
        profile,
64      preforms basic background correction, pushes data via TANGO into
        the ML model,
65      triggers inference in the TANGO device, retrieves prediction
        results and
66      publishes results via BLISS to be saved together with original
        scans in
67      the NeXus hdf5 file produced by the BLISS NeXusWriter
68      """
69
70      # specify the tango uri of the device server embedding the ML-Model
71      tango_uri_or_proxy = ("//VISA-HOST.esrf.fr:PORT/PATH/TO/TANGO-ML-
        DEVICE",)
72
73      # specify counters that should be used to collect the data from the
         scan object
74      TTH = setup_globals.gam  # gam: angle of detector axis
75      CROI = "pilatus300k:roi_counters:roi2_sum"  # ROI with specular
        signal
76      BGROI1 = (
77          "pilatus300k:roi_counters:roi3_sum"  # ROI left of CROI (half
        the size of CROI)
78      )
79      BGROI2 = (
80          "pilatus300k:roi_counters:roi4_sum"  # ROI right of CROI (half
        the size of CROI)
81      )
82      TRANSM = "autof_eh1:transm"  # transmission of Autoabsorber
83
84      # extract detector angle from scans
85      tth_data = np.concatenate(np.array([x.get_data(TTH) for x in scans
        ]))
86
87      # extract transmission of attenuators from scans
88      transm_data = np.concatenate(np.array([x.get_data(TRANSM) for x in
        scans]))
89
```

```python
# extract measured intensities from scans and perform basic
background correction
intens_data = np.concatenate(
    np.array(
        [x.get_data(CROI) - x.get_data(BGROI1) - x.get_data(BGROI2)
for x in scans]
    )
)

# establish communication with tango device that is linked to ML
model
ml_device = tango.DeviceProxy(tango_uri_or_proxy)

# deal with priors and preprocessing parameters
priors = priors or Priors()
preprocess = preprocess or PreprocessParams()

# push data to tango device
ml_device.input_tth = tth_data
ml_device.input_transm = transm_data
ml_device.input_intensity = np.clip(intens_data, 0, np.infty)
ml_device.input_count_time = [0]
ml_device.input_priors = priors.to_arr()
ml_device.postprocess_max_d_change = priors.max_d_change
ml_device.postprocess_fit_growth = priors.fit_growth
ml_device.preprocess_beam_width = preprocess.beam_width
ml_device.preprocess_sample_length = preprocess.sample_length
ml_device.preprocess_energy_keV = preprocess.energy_keV

# run inference (here built to run synchronous in blocking fashion)
ml_device.set_timeout_millis(5000)
ml_device.predict()

# retrieve predicted parameters
# polished parameters refers to results of subsequent LMS fit after
inference
polished_params = ml_device.prediction_parameters_polished
param_names = ml_device.prediction_parameter_names

# create summary metadata structure that is to be published via
BLISS
info = {"xrr": dict(zip(param_names, polished_params))}

# creates a new entry in the hdf5 file generated by BLISS and redis
DB for online access
seq = Sequence(
    title=f"AI_fit_thickness_{int(polished_params[0])}_Ang",
scan_info=info
)

# specify additional data entries that should appear under "
measurement" in the h5 file
seq.add_custom_channel(AcquisitionChannel("raw_tth", float, ()))
seq.add_custom_channel(AcquisitionChannel("corrected_intensity",
float, ()))
seq.add_custom_channel(AcquisitionChannel("raw_q", float, ()))
```

```python
137    seq.add_custom_channel(AcquisitionChannel("predicted_intensity",
       float, ()))
138    seq.add_custom_channel(AcquisitionChannel("predicted_q", float, ())
       )
139    seq.add_custom_channel(AcquisitionChannel("params", float, ()))
140    seq.add_custom_channel(AcquisitionChannel("params_names", str, ()))
141    seq.add_custom_channel(AcquisitionChannel("polished_intensity",
       float, ()))
142    seq.add_custom_channel(AcquisitionChannel("polished_params", float,
       ()))
143
144    # add references to the raw data scans as well as process data and
       analysis results
145    with seq.sequence_context() as scan_seq:
146        for scan in scans:
147            scan_seq.add(scan)
148
149        seq.custom_channels["raw_tth"].emit(tth_data)
150        seq.custom_channels["corrected_intensity"].emit(ml_device.
       prediction_refl)
151        seq.custom_channels["raw_q"].emit(ml_device.prediction_q)
152        seq.custom_channels["predicted_intensity"].emit(
153            ml_device.prediction_refl_predicted
154        )
155        seq.custom_channels["params"].emit(ml_device.
       prediction_parameters)
156        seq.custom_channels["params_names"].emit(ml_device.
       prediction_parameter_names)
157        seq.custom_channels["polished_intensity"].emit(
158            ml_device.prediction_refl_predicted_polished
159        )
160        seq.custom_channels["polished_params"].emit(polished_params)
161
162    # return polished_params e.g. for closed loop feedback
163    return polished_params
```

## SI-3. Asynchronous BLISS integration to receive events via the BLISS data streaming API

The code snippet below represents basically a consumer of events emitted via the event based data streaming api of BLISS and publishes produced results using a beacon channel.

References: `https://bliss.gitlab-pages.esrf.fr/bliss/1.10.x/data/data_redis_api_low.html` (accessed: 05.2023)

`https://bliss.gitlab-pages.esrf.fr/bliss/1.10.x/beacon_channels.html` (accessed: 05.2023)

```python
from bliss.data.node import get_session_node
from bliss.config.channels import import Channel

# create BLISS Channel with unique name in the scope of the beamline
# to publish predictions results via beacon / redis
last_film_thickness = Channel("last_film_thickness", default_value=None
    )


def listen_scans_of_session(session):
    """
    script that can run in any bliss process
    (independent oft the process producing the data & controlling the
    hardware)
    takes as input argument the name of the bliss session publishing
    the data
    to the beamline beacon server / redis
    """

    session_node = get_session_node(session)
    print(f"Listening to {session}")

    received_scans = []
    for event_type, node, event_data in session_node.walk_on_new_events
        (
            include_filter=["scan"]
        ):
            # this loop deals with the event based data streaming api of
    BLISS
            #
            if event_type == event_type.NEW_NODE:
                print(f"new scan", node.db_name)

            elif event_type == event_type.END_SCAN:
                print(node.db_name)

                params = node.info.get("fastfit")
                if params.get("xrr_scan_group_first", False):
                    received_scans = [node]
                    print("add scan")
                if params.get("xrr_scan_group_last", False):
                    received_scans.append(node)
                    print("add scan and process")
                    process_scans(received_scans)


def process_scans(scans):
    """
    Run ML inference
    """

    # missing here: code similar to synchronous version in example
    above
    # interacting with tango or triton server to retrieve prediction

    # ...
```

```
51
52     thickness = prediction["parameters_polished"][0]
53
54     # publish e.g. film thickness via bliss channel & beacon
55     # to make it available to all bliss processes running on the
       beamline
56     last_film_thickness.value = thickness
57
58     print("######## predicted thickness", thickness)
```

### SI-4. Postprocessing gradient-descent fit with optional growth rate parameter.

The functions below are used to perform a postprocessing polishing operation by fitting the measured curve via gradient descent initialized based on AI result. Optionally, it allows fitting the linear change of the thickness of the top layer as a function of $q$ for curves measured *in situ* during growth. Further details may be found in (Munteanu *et al.*, 2023).

References:

Munteanu, V., Starostin, V., Greco, A., Pithan, L., Gerlach, A., Hinderhofer, A., Kowarik, S. Schreiber, F. (2023). Neural network analysis of neutron and X-ray reflectivity data: incorporating prior knowledge for tackling the phase problem.

```
1  import numpy as np
2  from scipy.optimize import minimize, curve_fit
3
4  def abeles_np(
5          q: np.ndarray,
6          thickness: np.ndarray,
7          roughness: np.ndarray,
8          sld: np.ndarray):
9
10     c_dtype = np.complex128 if q.dtype is np.float64 else np.complex64
11
12     if q.ndim == thickness.ndim == roughness.ndim == sld.ndim == 1:
13         zero_batch = True
14     else:
15         zero_batch = False
16
```

```
17    thickness = np.atleast_2d(thickness)
18    roughness = np.atleast_2d(roughness)
19    sld = np.atleast_2d(sld)
20    batch_size, num_layers = thickness.shape
21    sld = np.concatenate([np.zeros((batch_size, 1)).astype(sld.dtype),
      sld], -1)[:, None]
22    thickness = np.concatenate([np.zeros((batch_size, 1)).astype(
      thickness.dtype), thickness], -1)[:, None]
23    roughness = roughness[:, None] ** 2
24
25    sld = sld * 1e-6 + 1e-30j
26
27    k_z0 = (q / 2).astype(c_dtype)
28    if len(k_z0.shape) == 1:
29        k_z0 = k_z0[None]
30    if len(k_z0.shape) == 2:
31        k_z0 = k_z0[..., None]
32    k_n = np.sqrt(k_z0 ** 2 - 4 * np.pi * sld)
33    k_n, k_np1 = k_n[..., :-1], k_n[..., 1:]
34
35    beta = 1j * thickness * k_n
36
37    exp_beta = np.exp(beta)
38    exp_m_beta = np.exp(-beta)
39    rn = (k_n - k_np1) / (k_n + k_np1) * np.exp(- 2 * k_n * k_np1 *
      roughness)
40
41    c_matrices = np.stack([
42        np.stack([exp_beta, rn * exp_m_beta], -1),
43        np.stack([rn * exp_beta, exp_m_beta], -1),
44    ], -1)
45    c_matrices = np.moveaxis(c_matrices, -3, 0)
46    m, c_matrices = c_matrices[0], c_matrices[1:]
47    for c in c_matrices:
48        m = m @ c
49
50    r = np.abs(m[..., 1, 0] / m[..., 0, 0]) ** 2
51    r = np.clip(r, None, 1.)
52
53    if zero_batch:
54        r = r[0]
55    return r
56
57
58  def standard_restore_params(fitted_params) -> dict:
59      num_layers = (fitted_params.size - 2) // 3
60      return dict(
61          thickness=fitted_params[:num_layers],
62          roughness=fitted_params[num_layers:2 * num_layers + 1],
63          sld=fitted_params[2 * num_layers + 1:],
64      )
65
66
67  def mse_loss(curve1, curve2):
68      return np.sum((curve1 - curve2) ** 2)
69
```

```python
def standard_refl_fit(
        q: np.ndarray, curve: np.ndarray,
        init_params: np.ndarray,
        bounds: np.ndarray = None,
        refl_generator=abeles_np,
        restore_params_func=standard_restore_params,
        scale_curve_func=np.log10,
        **kwargs):

    if bounds is not None:
        kwargs['bounds'] = bounds
        init_params = np.clip(init_params, *bounds)

    res = curve_fit(
        get_scaled_curve_func(
            refl_generator=refl_generator,
            restore_params_func=restore_params_func,
            scale_curve_func=scale_curve_func,
        ),
        q, scale_curve_func(curve),
        p0=init_params, **kwargs
    )

    curve = refl_generator(q, **restore_params_func(res[0]))
    return res[0], curve


def get_fit_with_growth(
        q: np.ndarray, curve: np.ndarray,
        init_params: np.ndarray,
        bounds: np.ndarray = None,
        init_d_change: float = 0.,
        max_d_change: float = 30.,
        scale_curve_func=np.log10,
        **kwargs):

    init_params = np.array(list(init_params) + [init_d_change])
    if bounds is not None:
        bounds = np.concatenate([bounds, np.array([0, max_d_change])
[..., None]], -1)

    params, curve = standard_refl_fit(
        q, curve, init_params, bounds, refl_generator=
refl_with_changing_params,
        restore_params_func=get_restore_params_with_growth_func(q_size=
q.size, d_idx=0),
        scale_curve_func=scale_curve_func, **kwargs
    )
    params[0] += params[-1] / 2
    return params, curve


def fit_refl_curve(q: np.ndarray, curve: np.ndarray,
                   init_params: np.ndarray,
                   bounds: np.ndarray = None,
```

```python
123                    refl_generator=abeles_np,
124                    restore_params_func=standard_restore_params,
125                    scale_curve_func=np.log10,
126                    **kwargs
127                    ) -> np.ndarray

128
129    fitting_func = get_fitting_func(
130        q=q, curve=curve,
131        refl_generator=refl_generator,
132        restore_params_func=restore_params_func,
133        scale_curve_func=scale_curve_func,
134    )

135
136    res = minimize(fitting_func, init_params, bounds=bounds, **kwargs)

137
138    if not res.success:
139        import warnings
140        warnings.warn(f"Minimization did not converge.")
141    return res.x

142
143
144 def get_scaled_curve_func(
145        refl_generator=abeles_np,
146        restore_params_func=standard_restore_params,
147        scale_curve_func=np.log10):

148
149    def scaled_curve_func(q, *fitted_params):
150        fitted_params = restore_params_func(np.asarray(fitted_params))
151        fitted_curve = refl_generator(q, **fitted_params)
152        scaled_curve = scale_curve_func(fitted_curve)
153        return scaled_curve

154
155    return scaled_curve_func

156
157
158 def get_fitting_func(
159        q: np.ndarray,
160        curve: np.ndarray,
161        refl_generator=abeles_np,
162        restore_params_func=standard_restore_params,
163        scale_curve_func=np.log10,
164        loss_func=mse_loss):

165
166    scaled_curve = scale_curve_func(curve)

167
168    def fitting_func(fitted_params):
169        fitted_params = restore_params_func(fitted_params)
170        fitted_curve = refl_generator(q, **fitted_params)
171        loss = loss_func(scale_curve_func(fitted_curve), scaled_curve)
172        return loss

173
174    return fitting_func

175
176
177 def restore_masked_params(fixed_params, fixed_mask):
178    def restore_params(fitted_params) -> dict:
```

```python
179        params = np.empty_like(fixed_mask).astype(fitted_params.dtype)
180        params[fixed_mask] = fixed_params
181        params[~fixed_mask] = fitted_params
182        return standard_restore_params(params)
183
184    return restore_params
185
186
187 def base_params2growth(base_params: dict, d_shift: np.ndarray, d_idx:
     int = 0) -> dict:
188    d_init = base_params['thickness'][None]
189    q_size = d_shift.size
190    d = d_init.repeat(q_size, 0)
191    d[:, d_idx] = d[:, d_idx] + d_shift
192
193    roughness = np.broadcast_to(base_params['roughness'][None], (q_size
     , base_params['roughness'].size))
194    sld = np.broadcast_to(base_params['sld'][None], (q_size,
     base_params['sld'].size))
195
196    return {
197        'thickness': d,
198        'roughness': roughness,
199        'sld': sld,
200    }
201
202
203 def get_restore_params_with_growth_func(q_size: int, d_idx: int = 0):
204    def restore_params_with_growth(fitted_params) -> dict:
205        fitted_params, delta_d = fitted_params[:-1], fitted_params[-1]
206        base_params = standard_restore_params(fitted_params)
207        d_shift = np.linspace(0, delta_d, q_size)
208        return base_params2growth(base_params, d_shift, d_idx)
209
210    return restore_params_with_growth
211
212
213 def refl_with_changing_params(q: np.ndarray, **kwargs):
214    return abeles_np(q[..., None], **kwargs).flatten()
```



\end{document}